# Electromagnetic Properties of a Hot and Dense Medium

**Samina Masood**[1]

Department of Physical and Applied Sciences, CSE,
Address, University of Houston Clear Lake, Houston TX 77058
*E-mail: masood@uhcl.edu*

We study the properties of an electromagnetically interacting medium in the presence of high concentration of electrons at extremely high temperatures and chemical potentials. We show that the electromagnetic properties of a medium such as the electric permittivity, magnetic permeability, magnetic moments and the propagation speed of electromagnetic waves as well as the corresponding particle processes depend on temperature and density of the medium. Electromagnetic properties of neutrinos are significantly modified due to their interactions with electrons when they propagate through such a medium of hot and dense electrons.

## 1. Introduction

We study the propagation of electromagnetic waves in a medium of electrons at the extreme conditions of temperature and densities. High temperature with ignorable values of chemical potential were available in the very early universe, whereas, extremely large chemical potential with large temperatures of the order of millions of kelvins is indicated in the cores of super dense stars. Large chemical potential and large magnetic fields are associated with high densities only. We use the renormalization scheme of QED (Quantum Electrodynamics) [1-3] to study the vacuum polarization tensor of photon as compared to the previously used effective potential approach [4]. All the components of vacuum polarization tensor of photon at high temperatures and densities of a medium are calculated in detail to study the propagation of light in different directions of medium. Our hypothetical system correspond to a heat bath of electrons at a very high equilibrium temperature and even larger density of electrons. Such type of systems can be identified as classical systems [5] because the chemical potential is large enough to overcome the thermal contribution. The standard form of vacuum polarization tensor in QED at finite temperature and density (FTD) can be written by replacing the electron propagator in vacuum by the statistically corrected propagator in real-time formalism, including the fermion distribution function $n_F$ in the usual notation as:

$$S(p) = \left[\frac{1}{p^2 - m^2} + \Gamma_F(p,\mu)\right] \text{ with } \Gamma_F(p,\mu) = 2\pi i \delta(p^2 - m^2)[\theta(p_0)n_F(p,\mu)$$

The complete vacuum polarization tensor $\pi_{\mu\nu}(K)$ can be written in the form of longitudinal $\pi_L(K)$ and transverse $\pi_T(K)$ components as given in Ref. [1]

### 2 Calculation of Electric Permittivity and Magnetic Permeability

$\pi_L(K)$ and $\pi_T(K)$ can be evaluated from $\pi_{\mu\nu}(K)$ using the renormalization scheme [1-3] of QED at finite temperature and density. The electric permittivity $\varepsilon(k)$ and the magnetic permeability $\mu(k)$ can be expressed in terms of the longitudinal ($\pi_L$) and transverse component ($\pi_T$) of the polarization tensor as

$$\varepsilon(k) = 1 - \frac{\pi_L(K)}{k^2}, \qquad \frac{1}{\mu(k)} = \frac{k^2\pi_T(K) - \omega^2\pi_L(K)}{k^2 K^2},$$

whereas, the electric permittivity in a hot and dense medium has been calculated as:

$$\varepsilon(K) \cong 1 - \frac{4e^2}{\pi^2 K^2}\left(1 - \frac{\omega^2}{\mathbf{k}^2}\right)\left\{\left(1 - \frac{\omega}{2\mathbf{k}}\ln\frac{\omega+\mathbf{k}}{\omega-\mathbf{k}}\right)\left(\frac{ma(m\beta,\mu)}{\beta} - \frac{c(m\beta,\mu)}{\beta^2}\right)\right.$$
$$\left. + \frac{1}{4}\left(2m^2 - \omega^2 + \frac{11k^2 + 37\omega^2}{72}\right)b(m\beta,\mu)\right\},$$

and the magnetic permeability as:

$$\frac{1}{\mu(K)} \cong 1 - \frac{2e^2}{\pi^2 k^2 K^2}[\omega^2\left\{\left(1 - \frac{\omega^2}{\mathbf{k}^4} - \left(1 + \frac{\mathbf{k}^2}{\omega^2}\right)\left(1 - \frac{\omega^2}{\mathbf{k}^2}\right)\frac{\omega}{2\mathbf{k}}\ln\frac{\omega+\mathbf{k}}{\omega-\mathbf{k}}\right\}\right.$$
$$\left. \times\left(\frac{ma(m\beta,\mu)}{\beta} - \frac{c(m\beta,\mu)}{\beta^2}\right) - \frac{1}{8}\left(6m^2 - \omega^2 + \frac{129\omega^2 - 109k^2}{72}\right)b(m\beta,\mu)\right\}].$$

with the correct definition of Masood's abc functions [1-3] in a hot and dense medium. We evaluate electric permittivity and the magnetic permeability for extreme temperatures of the early universe with ignorable density in the first case and then we also evaluate them for extremely dense systems with negligible temperatures as compared to the corresponding chemical potentials.

Case I: At high temperature $T \gg \mu$

For $\omega \gg k$ the longitudinal and transverse components of the vacuum polarization tensor can be evaluated as

$$\pi_L = -\frac{\omega^2 e^2 T^2}{3k^2}, \qquad \pi_T = \frac{\omega^2 e^2 T^2}{6k^2}, \qquad \varepsilon(k) = 1 + \frac{\omega^2 e^2 T^2}{3k^4}, \qquad \frac{1}{\mu(k)} \cong \frac{\omega^4 e^2 T^2}{3k^4 K^2},$$

For $\omega \ll k$ the corresponding longitudinal and transverse components of the vacuum polarization tensor will look like

$$\pi_T = \frac{e^2 T^2}{6}, \qquad \pi_L = \frac{e^2 T^2}{3} \qquad \varepsilon(k) = 1 - \frac{e^2 T^2}{6k^2}, \qquad \frac{1}{\mu(k)} \cong \frac{\omega^4 e^2 T^2}{3k^4 K^2},$$

## 3. Results and Conclusions:

It is interesting to note $T^2$ dependence at extremely high temperatures and $\mu^2$ dependence at extremely high chemical potentials. Electromagnetic properties of a medium are modified at high temperatures and chemical potential. In the extremely dense systems, the electric permittivity and the magnetic permeability change as a functions of chemical potential, the wave number k and the frequency omega are expressed in units of electron mass. For any given value of temperature and chemical potential, they depend on the wavenumber and the frequency of the propagating photon. In this paper, only the extreme situations are considered to find the dependence of electric permittivity and the magnetic permeability on the wavenumber of photon. Fig. 1.shows a distinct behavior of electric permittivity and shows a slight dip in permittivity at the minimum before it attains a constant value and does not depend on the wavenumber any more.Plots of Electric permittivity (red), magnetic permability (blue) and the propagation speed (Grey) for another limit is given as a function of wave number in Fig.2. More details can be seen in some of the recent works [6], submitted for publication.

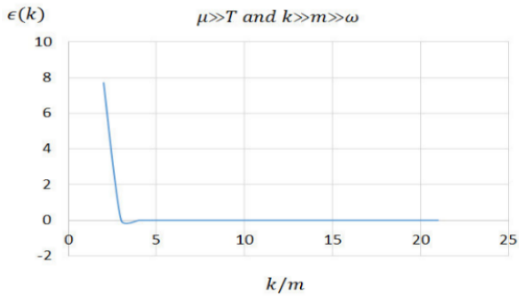
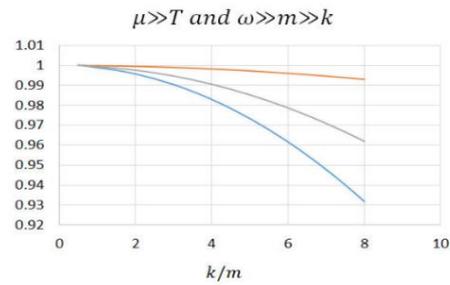

Fig.1: Electric permittivity is plotted as a function of the photon wavenumber k in units of m as hk/2π=mc²

Fig, 2: Magnetic permeability is plotted in red and electric permittivity in blue as a function of k in units of m using hk/2π=mc²